# On spatial electron-photon entanglement

Eitan Kazakevich, Hadar Aharon and Ofer Kfir*

[1]School of Electrical Engineering, The Iby and Aladar Fleischman Faculty of Engineering, Tel Aviv University, Tel Aviv, 69978
(*)correspondence email: kfir@tauex.tau.ac.il

*Abstract*

Free electron beams and their quantum coupling with photons is attracting a rising interest due to the basic questions it addresses and the cutting-edge technology these particles are involved in, such as microscopy, spectroscopy, and quantum computation. This work investigates theoretically the concept of electron-photon coupling in the spatial domain. Their interaction is discussed as a thought experiment of spontaneous photon emission from a dual-path free-electron (free e⁻) beam. We discuss a retro-causal paradox that may emerge from naively extending perceptions of single-path e⁻-photon coupling to transversely separated paths, and its resolution through the physics of two-particle interference. The precise spatial control of electrons and photons within e⁻-microscopes enables manipulation of their respective states, thus, such instruments can harness position-encoded free-e⁻ qubits for novel quantum sensing and the transfer of quantum information.

The phenomenon of entanglement captured the imagination of physicists since its conception, and now, through the advent of quantum computation and adjacent technologies, it fascinates the broader public as well. Upon its conception entanglement was used to contest the validity of quantum mechanics by apparent paradoxes it raises. Famous examples were proposed by Einstein, Podolsky, and Rosen (EPR) [1], and the discussion by Wheeler, Scully and Drühl's on quantum erasers [2,3]. These paradoxes sharpen one's understanding of the quantum nature of entangled particles with respect to locality, realism, and causality. Being a general phenomenon, entanglement is mandatory for quantum computation hardware such as atoms [4,5], ions [6,7], and condensed matter [8,9], as well as for photons [10], which are currently the *only suitable entity for quantum communication* at any scale [11,12].

As free photons, free electrons are fundamental particles – the simplest fermions within the standard model of particle physics. Electron microscopes routinely form high quality $e^-$-beams and control their parameters such as the kinetic energy, beam size and convergence angles, and integrate light insertion and collection apparatuses [13–19]. Even with such a substantial control of a fundamental particle, the concept of entanglement between free electrons and other fundamental entities, such as photons, has only been conceived recently [20,21]. Thus, free electrons may potentially mediate quantum-information transfer between, say, a photon and other quantum systems. Since modern $e^-$-microscopy technology can focus relativistic $e^-$-beams to below an angstrom [22], schemes that entangle the electrons could be selective to atomic-scale targets and surpass the diffraction limit of quantum light. Thus, an elaborate effort is made to optimize photonic structures for maximizing and utilizing electron-photon coupling [13,23–27], and to estimate its ultimate theoretically its supremum across the electromagnetic spectrum [28,29].

Much of the experimental work in electron-photon coupling was on stimulated phenomena, executed by the interaction of classical fields with the electron wavefunction, such as PINEM (photon-induced nearfield electron microscopy) [30–33] and ponderomotive modulation [34–37]. Proving a powerful nanophotonic tool, PINEM enabled nanoscopic imaging of optical modes [14,30,31,38–46], control the electronic wavefunction [15,31,47–49], formation of attosecond $e^-$-pulses [48,50,51] and imaging of light-driven dynamics with sub-cycle precision [16,49]. The quantum-fields description of PINEM is of an unitary electron-photon scattering operator [20,46] acting on a photonic coherent-state and stimulating energy transitions of the electron. The PINEM $e^-$-spectrum emerges when a vanishingly small quantum coupling is compensated for by the strong stimulus of a highly populated coherent state [20]. The spin of the electron typically contributes weakly to the interaction with light or between electrons [52] in the beam and hence we consider it an unperturbed and negligible degree of freedom.

Spontaneous electron-photon coupling, such as observed through EELS (electron energy-loss spectroscopy) [53–55], CL (cathodoluminescence) [17,56,57] or both [13,26], may be considered as more quantum since they address exchanges of single quanta. In high-resolution electron microscopes EELS and CL probe spatial quantum properties of the density of states since the photonic modes are initially at their vacuum level [20,55,58–60]. Free electrons couple spontaneously to a variety of polaritons through their electric field [13,26,54,55,61–68]. In a series of publications by Potapov, Verbeeck and co-workers, a transversely separated bi-partite electron wavefunction was probing the coupling to plasmons and intramolecular polaritons [69–71]. These experiments probed only the electrons, analyzing a loss of interference visibility in the far-field, thus, tracing out the inaccessible excitation degree of freedom. Complementary, Remez et al, investigated the plasmon-mediated photon emission from a spatially extended e⁻-wavefunction, finding that the electron's transverse coherence had no effect on the CL [72]. The role of the plasmons in both cases was quantitative as they offer a high probability of energy exchange with the electron, up to a few %, due to the macroscopic response of electrons at the metallic conduction band [54,61]. However, pure photons are profoundly more important from the perspective of quantum technologies. They can carry quantum information about the interaction towards a distant detector, where number state measurements [20,26] or quadrature measurements through homodyne detection [16,73] can be employed. Hence, photons rather than other various polaritons are key to novel quantum technologies with free-electron beams.

The analytic treatment of entanglement between electrons and photons referred, to date, to an approximately one-dimensional path of the e⁻-beam, which is a good approximation due to the beam's extreme transverse confinement with respect to the optical wavelength. The emphasis on the propagation direction is not accidental. A non-vanishing e⁻-photon coupling needs a long (μm scale) momentum-matched and uniform interaction, which further highlight the importance of the line-like electron path. However, reaching maximally entangled e⁻-photon states through one dimensional interaction necessitates strong coupling amplitudes, which remain elusive [20,24,26,27]. Furthermore, even when strong coupling is reached, electron energy-ladder quantum states will be challenging substrate for realizing qubits due to their essentially infinite energy extent [25,74].

Here, we expand the concept of quantum electron-photon coupling to the spatial domain. We pose the problem as a thought experiment that maps the two slits to electron- and photon qubits. The spatial photonic states that are propagating within well-separated waveguides and measured by single-photon detectors directly or after a beam-splitter. The electrons are described similarly. To emphasize the novel features, we start by relying on common understandings of a one-dimensional electron path and show that when extended naively to the transverse dimension it may lead to a causality paradox. The paradox is resolved by handling the detection of the electron-photon hybrid as a "two-particle interference", as defined by Jaeger [75]. Importantly, Jaeger's approach explains the experiments of Potapov et al., where spatially shaped e⁻-wavefunction coupled to polaritons [70–72]. We show explicitly that separated e⁻-photon interactions can form maximally entangled, Bell-like, pairs of an electron and a photon. Thus, this work opens a path for harnessing free-electrons as quantum-information carriers, at arbitrarily low coupling amplitudes to photons.

*The thought experiment – single e⁻-path*

To properly refer to the current state-of-the art understanding of electron photon coupling, in this section we examine the properties of the electron-photon coupling in a simple geometry with a systematic disparity: the electron travels along a single line whereas the photon modes travel in two single-mode waveguides parallel to it, over some interaction distance (See Fig. 1A). The photons are assumed to be generated coherently, having their phase temporally locked to the electron timing [73]. A matching between the velocities of the electron and the optical phase in the waveguide could enable an efficient photon generation over an extended interaction length [15,20,73,76]. As illustrated in Fig. 1A, we assume that the waveguides are far from the electron path and from each other until they enter smoothly into the interaction region, where they have a fixed distance. Passing to the left- and right-hand sides of the electron, we mark n-photons states in the waveguides as $\left|n^{ph}_{left\,/\,right}\right\rangle$, respectively, at an implicit frequency $\omega$. Their separation downstream marks the end of the interaction region, from which they are guided a symmetric beam splitter followed by photon counters marked as D1 and D2. The corresponding states are marked $\left|n^{ph}_{D1\,/\,D2}\right\rangle$, respectively. We allow a tunable degree of freedom, $\phi$, for relative phase between the two inputs before entering the beam splitter.

Consider an electron path that within the interaction region is parallel and equidistant from the two optical waveguides on its left-hand and right-hand sides. If the waveguides are identical, the complex coupling amplitude, $g_{Qu}$, to each of the optical modes is equal since it is a geometrical factor. Prior to the interaction the combined electron-photon system can be marked by separable states of an electron at its zero-loss energy, $|E_0^e\rangle$ and no photons,

$$|\psi_1\rangle = |E_0^e\rangle \left|0^{ph}_{left}, 0^{ph}_{right}\right\rangle. \tag{1}$$

Assuming the coupling is weak, $|g_{Qu}| \ll 1$, the state after interaction is

$$|\psi_2\rangle \approx |\psi_1\rangle + g_{Qu}|E_0 - \hbar\omega\rangle \left(\left|1^{ph}_{left}, 0^{ph}_{right}\right\rangle + \left|0^{ph}_{left}, 1^{ph}_{right}\right\rangle\right). \tag{2}$$

The perturbation added to the initial $|\psi_1\rangle$ is a state in which the electron energy lost a photon-energy quantum, $\hbar\omega$, for populating the photon field at the left or right waveguide. $\hbar$ is the reduced Plank's constant.

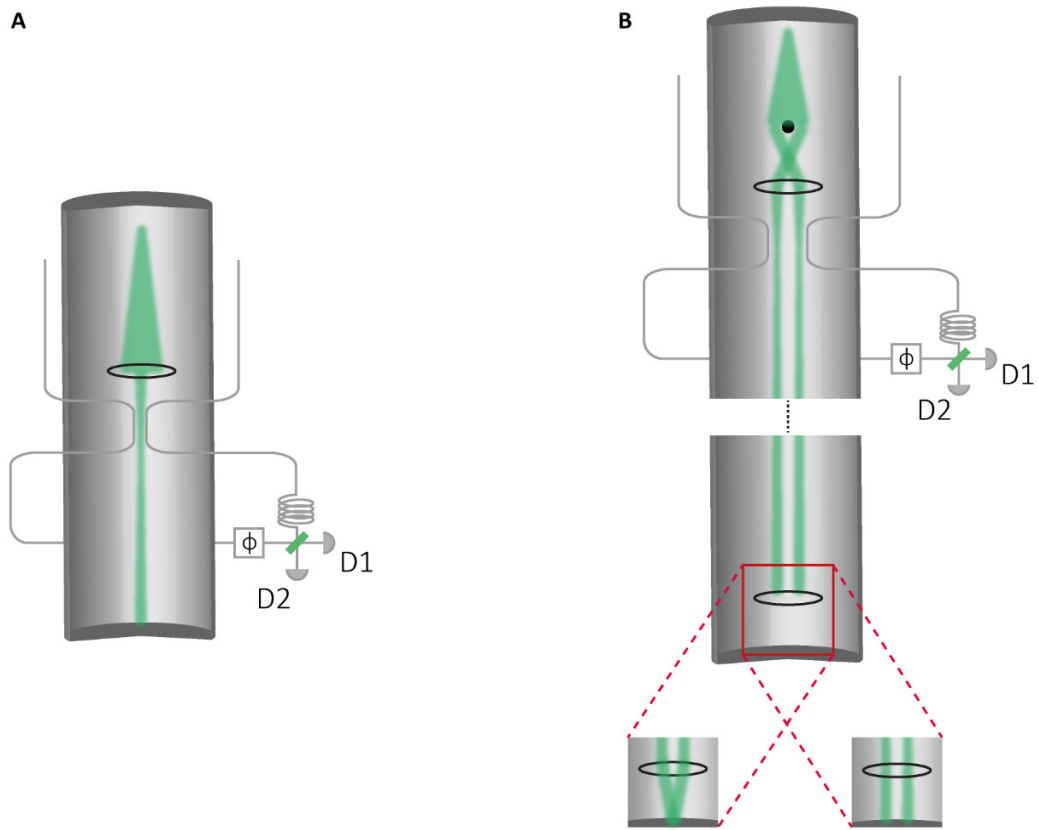

FIG. 1 – Scheme of the proposed experiments for single- vs. double path of free-electron beams. (a) Single path electron coupled to two identical optical waveguides that acquire a relative phase $\phi$ prior to their mixing in a beam splitter. $\phi$ can be set to exclude CL photons from arrival to one of the detectors, say D2. (B) When the electron beam is in a spatial two-path superposition both electron and photon states can be controlled: the electrons paths can be either mixed (left inset) or be distinguished (right inset). The photons can be detected either with or without a beam splitter. Naive drawing of parallels from the single-e⁻-path to the double path results in a seeming paradox.

As the two photons may acquire a different phase as they propagate to the beam splitter, the phase difference depends only on the geometry, e.g., the length of each of the respective fibers. Without a loss of generality, we can assume a setting in which the photons exit at the port D1. One can tune the balance between D1 and D2 by tuning the optical retardation phase of one of the fibers, marked as $\phi$,

$$P(D1) = \frac{1}{2}(1 - sin\phi), P(D2) = \frac{1}{2}(1 + sin\phi).$$

Thus, we get interference between the photonic paths. The double fiber acts as an electron-photon-coupling analogue to Young's double slit experiment, similar to Feynman's famous example [77], but in these famous experiments each single photon is split into a superposition of two paths, here the electron creates them in this superposition directly.

*Thought experiment – dual e⁻-path*

One can extend the above logic to a dual electron path such that within the interaction region it is in a spatial super-position of two transversely shifted paths, at $x = x_0$ and $x = -x_0$, which we term left and right. Each of these paths is parallel to a single-mode optical waveguide in its vicinity (See Fig. 1B). We assume that each e⁻-path has a complex coupling amplitude $g_{Qu}$ to the adjacent fiber and a negligible coupling amplitude to the other fiber, where $g_{Qu}^{left} = g_{Qu}^{right} = g_{Qu}$. The fibers enter an ideal symmetric beam splitter and detectors D1 and D2, as before.

Considering a transmission e⁻-microscope apparatus means that the measurement of the electron paths can be made either distinguishable or indistinguishable, controllably. The path of the electron can be determined by imaging on the e⁻-camera the focal plane of the e⁻-beams. Alternatively, if both e⁻-paths are projected over the camera with full overlap the left- or right interaction trajectories are indistinguishable. Thus, the user can toggle the indistinguishability of the electron paths to "on" or "off" using the post-sample e⁻-optics. Similarly, the overlap between the two electron beams on the camera can be tuned continuously from zero ("off") and up to 100% ("on") by a combination of electron lenses and biprisms.

One may expect that for a certain tuning of the phase $\phi$ the photons end up systematically in the detector D1, as in the previous section. However, since a deterministic detection at D1 requires the suppression of the which-path information, a consistent emission into D1 can occur only if the electron optics are in an "on" state. Hence the paradox that the naïve assumptions lead to. They seemingly predict the manipulation of a particle's probability distribution (the photon) by controlling another (the electron) [75]. In other words, the paradox emerges from the assumption of a deterministic photon phase, $\phi$. More explicitly, when the electron's path indistinguishability is set to "off", the interaction path is measured either at the left *or* the right side, so the photon hits D1 and D2 randomly with equal probability, regardless of $\phi$. At an "on" state the photon is in a superposition of the two paths, and if the phase between them is deterministic the photons can be set to hit D1 consistently. The "on" and "off" states settings of the electron optics seem to change the statistics of the photon detection. Phrasing the paradox as a space-time impossibility, such an effect suggests that if the electrons are sent to a distant observer, he/she would determine the statistics of the photon detection placed near the interaction point, in a retro-causal manner.

*Explaining the paradox*

The paradox is resolved by adding the spatial entanglement of the two paths, and drawing parallels to other quantum communication protocols, such as quantum key distribution and quantum eraser [2,3,11,78–80]. As we show below rigorously, the detection is always balanced, with photons hitting D1 and D2 randomly, *regardless* of the manipulations of the electron state made after the interaction, or the chosen optical retardation between the two detectors, $\phi$. Quantum information transfer in this system would emerge from spatial correlations between the electrons and co-incident photons.

To analyze the electron-photon system, we follow a similar approach as above (eq. 2) and calculate the electron properties, conditional on photon measurement at D1 or D2. The state prior to the interaction is

$$|\Psi_1\rangle = \tfrac{1}{\sqrt{2}}(|E_0^e, left\rangle + |E_0^e, right\rangle)|0_{left}^{ph}, 0_{right}^{ph}\rangle, \tag{3}$$

Where now we include the two paths of the electron with its initial energy, marked left and right, as different states. In the weak coupling regime, the state after interaction is,

$$|\Psi_2\rangle \approx |\Psi_1\rangle + g_{Qu}\left(|E_0^e - \hbar\omega, left\rangle|1_{left}^{ph}, 0_{right}^{ph}\rangle + |E_0^e - \hbar\omega, right\rangle|0_{left}^{ph}, 1_{right}^{ph}\rangle\right). \tag{4}$$

After the beam-splitter, the photon-path subscript index is properly written according to the detector it will be hitting, D1 or D2. We can arbitrarily choose the phase delay and the beam splitter matrix such that they jointly act as a Hadamard gate [81], $\begin{pmatrix}\hat{a}_{D1}\\\hat{a}_{D2}\end{pmatrix} = \tfrac{1}{\sqrt{2}}\begin{pmatrix}1 & 1\\1 & -1\end{pmatrix}\begin{pmatrix}\hat{a}_{left}\\\hat{a}_{right}\end{pmatrix}$, with $\hat{a}_i$ a photonic annihilation operator at frequency $\omega$ and $i$ is an index for $D1, D2, left, right$. The state after the beam splitter is then,

$$|\Psi_3\rangle \approx |\Psi_1\rangle + g_{Qu}\tfrac{1}{\sqrt{2}}\begin{pmatrix}|E_{-1}^e, left\rangle(|1_{D1}^{ph}, 0_{D2}^{ph}\rangle + |0_{D1}^{ph}, 1_{D2}^{ph}\rangle)\\+ |E_{-1}^e, right\rangle(|1_{D1}^{ph}, 0_{D2}^{ph}\rangle - |0_{D1}^{ph}, 1_{D2}^{ph}\rangle)\end{pmatrix}, \tag{5}$$

and collecting the terms according to D1 and D2, brings it to the form,

$$|\Psi_3\rangle \approx |\Psi_1\rangle + g_{Qu}\tfrac{1}{\sqrt{2}}\begin{pmatrix}|1_{D1}^{ph}, 0_{D2}^{ph}\rangle(|E_{-1}^e, left\rangle + |E_{-1}^e, right\rangle)\\+ |0_{D1}^{ph}, 1_{D2}^{ph}\rangle(|E_{-1}^e, left\rangle - |E_{-1}^e, right\rangle)\end{pmatrix}. \tag{6}$$

We abbreviate the e⁻-energy $E_0^e - \hbar\omega$, as $E_{-1}^e$. Here, the spatial entanglement between the electron and the photon is clear. The phase between the left and right paths of the electron's wave-function correlate with the photon detection: $\frac{\pi}{2}$ for hits at D1 and $-\frac{\pi}{2}$ for D2. To observe these correlations, one needs to detect the electrons in states that are sensitive to such a relative phase. Had we had a state mixer for free electrons which is equivalent to a photonic beam-splitter, we could have used it and get the electrons in final states $|e_+\rangle$ and $|e_-\rangle$. In such a hypothetical electron-photon state,

$$Hypothetic: g_{Qu}(|1_{D1}^{ph}, 0_{D2}^{ph}\rangle|e_+\rangle + |0_{D1}^{ph}, 1_{D2}^{ph}\rangle|e_-\rangle),$$

The hypothetical states allow us to describe the e⁻-photon entanglement symmetrically. When the measurements of the two systems share basis, they are fully correlated. If the beamsplitter is used for both, detections at D1 and D2 coincide with hypothetical hits on the detectors for $|e_+\rangle$ and $|e_-\rangle$, respectively. Similarly, the e⁻-state is imaged in real space and the photonic beam splitter is removed, both systems measure in the left/right basis, where they correlate, according to eq. (4). This resolves the paradox. Detection at the left or right paths is a random process with equal probability for both the electron and the photon, as is the detection at D1 or D1 (or the hypothetical $e_+/e_-$), where only co-incident measurements reveal correlations. Referring to the case of a distant measurement of the electrons, the comparison of these measurement results over a classical communication channel solves the causality issue.

We now turn to describe how the spatial electron-photon entanglement can be expressed in a realistic experimental instrument in an electron microscope, without the hypothetic binary electron $e_{+/-}$ states (Illustrated in Fig. 2). The photon system does not require any additional treatment. Photons emanating from waveguides interacting with grazing electron beams were measured by Feist et al [26], including their correlation with the electron's energy loss. Thus, we focus on the electron sub-system, envisioning a unitary operator $\widehat{U}_i$ that governs the indistinguishability between left and right paths. It combines the effects of free-space propagation and electron optics such as lenses, deflectors, and bi-prisms. The subscript 'i' will be used to quantify the e⁻-path erasure. Assuming $\widehat{U}_i$ represents the electron propagation up to a detector (e.g., a camera) the final electron-photon state becomes,

$$|\Psi_4\rangle = \widehat{U}_i^e|\Psi_3\rangle. \tag{7}$$

Therefore, the photon detection probabilities at D1, D2 are:

$$\langle\Psi_4|\hat{n}_{D1}|\Psi_4\rangle = |g_{Qu}|^2 \frac{1}{2}(\langle E_{-1}^e, left| + \langle E_{-1}^e, right|)\widehat{U}_i^{e\dagger}\widehat{U}_i^e(|E_{-1}^e, left\rangle + |E_{-1}^e, right\rangle) \tag{8}$$

and

$$\langle\Psi_4|\hat{n}_{D2}|\Psi_4\rangle = |g_{Qu}|^2 \frac{1}{2}(\langle E_{-1}^e, left| - \langle E_{-1}^e, right|)\widehat{U}_i^{e\dagger}\widehat{U}_i^e(|E_{-1}^e, left\rangle - |E_{-1}^e, right\rangle). \tag{9}$$

$\widehat{U}_i$ is quite general. It can maintain the electron which-path information or erase it, even partially. It is helpful to consider two distinct and realistic cases for $\widehat{U}_i^e$: one, where it represents an imaging of the interaction region, mapping a point $(x,y)$ of the wave-function at the interaction region, $\psi(x,y)$, to a magnified replica $\psi(ax_{cam}, ay_{cam})e^{i\chi(x,y)}$, where $\frac{1}{a}$ is the magnification factor and $\chi(x,y)$ is a phase profile that can be omitted when measuring the probability, $|\psi(ax_{cam}, ay_{cam})|^2$. Alternatively, $\hat{U}_i$ can project the far-field on the camera, $\mathcal{F}[\psi(x,y)]_{\{k_x,k_y\}}$, where $\mathcal{F}$ is the Fourier transform and $k_x, k_y$ are the spatial frequencies. The position of a spatial frequency component on the camera is $(x_{cam}, y_{cam}) = (k_x, k_y)\lambda_{DB}f_l$, where $\lambda_{DB}$ is the De-Broglie wavelength and $f_l$ is the effective focal length of the electron optical system. In the simple two-path scenario we address in this paper the electron wavefunction is approximated as two spatial delta-functions on the x-axis, spaced by $2x_0$, where $|E^e_{-1}, left\rangle = \delta(y)\delta(x-x_0)$ and $|E^e_{-1}, right\rangle = \delta(y)\delta(x+x_0)$. When set to imaging we approximate $\hat{U}_i$ as the unity operator, omitting its magnification. The probabilities for photon detection at D1 or D1 (eq. (8)-(9)) are equal and the phase between the left and right paths is meaningless. When the microscope maps the far-field on the camera, $\hat{U}^e_i$ is proportional to a Fourier transform. According to eq. (8) and (9), a detection at D1 will be fully correlated with a squared cosine electron probability pattern along the $x_{cam}$ axis, and a detection at D2 would correlate with a sinusoidal one.

$$\langle\Psi_4|\hat{n}_{D1}|\Psi_4\rangle = |g_{Qu}|^2 \cos^2(2x_0 k_x) = |g_{Qu}|^2 \cos^2\left(2x_0 \frac{x_{cam}}{\lambda_{DB}f_l}\right), \qquad (10)$$

$$\langle\Psi_4|\hat{n}_{D2}|\Psi_4\rangle = |g_{Qu}|^2 \sin^2(2x_0 k_x) = |g_{Qu}|^2 \sin^2\left(2x_0 \frac{x_{cam}}{\lambda_{DB}f_l}\right). \qquad (11)$$

Thus, in this concrete example of the paradox resolution, photon detection fully correlates with the shift of a sinusoidal electron pattern. Detection co-incident with clicks at D1 are shifted by half-a period from those co-incidents with D2. If one ignores the photonic degrees of freedom and traces them out, the electron state forms a uniform probability "blob" in the far field or random hits on the magnified image of the left/right paths. Considering eq. (8)-(11) through the perspective of the paradox, one sees that regardless of a particular choice of $\hat{U}_i$, the incidence probability at D1 and D2 is equal, carrying no a-priori information on the electron path.

The entanglement of spatially varying free-electron states and the photons they emit through coupling with a dielectric structure is analogous to the general problem of two particle interference. From a formalism perspective it must be so, since the mathematical description is identical, so are the results. For example, one can map our conceptual experiment to Bell-pair photon generation by spontaneous parametric down-conversion (SPDC) [11,82]. In down-conversion, the pump photon is annihilated, and two lower-energy photons emerge, with a single quantum populating in each of their mode. The electron quantum system may seem different since the electron does not vanish; however, its initial-energy state does since the e⁻ is no longer populating the energy state $E_0^e$. The energy is conserved by the excitation of a state at the reduced e⁻-energy, $E_{-1}^e$, alongside a photon.

The electron-photon entanglement can be viewed from the perspective of the spatial symmetry. The physical system is mirror-symmetric for $x \to -x$, including the waveguides and the electron path, hence, under such mirroring the system can have only eigenvalues $\pm 1$. We argue that the mirroring eigenvalue of the final e⁻-photon state is equal to that of the initial state, i.e., to the split-beam electron wavefunction. Assuming a symmetric input (e.g., eq. (3)), the final electron and photon states are either both positive or both negative, symmetric, or anti-symmetric, respectively. Since $\phi$ can be tuned such that the beam splitter sends a parity-symmetric photon to D1 and an asymmetric photon to D2 these coincide with a final symmetric e⁻-wavefunction $\cos(2x_0 k_x)$ or an anti-symmetric $\sin(2x_0 k_x)$, respectively. The $\pm 1$ eigenvalues of mirroring over the YZ plane serves as an analogue to standard binary measurement bases, such as polarization. Once these parallels are clearly drawn one can reconcile any claim for paradoxical superluminal communication or similar. Importantly, it also means that applications of photonic Bell pairs can be extended to electron-photon pairs. In contrast to purely photonic Bell-pairs, electron-photon pairs combine complementary properties: photons can carry quantum information over distances since they are weakly interacting, and electrons can probe with specificity well below the size of a single atom. Hence, since quantum-optical technology is mature and electron-optics is precise at the atomic scale, electrons could mediate quantum information between the macroscopic world and individual quantum systems with arbitrary precision.

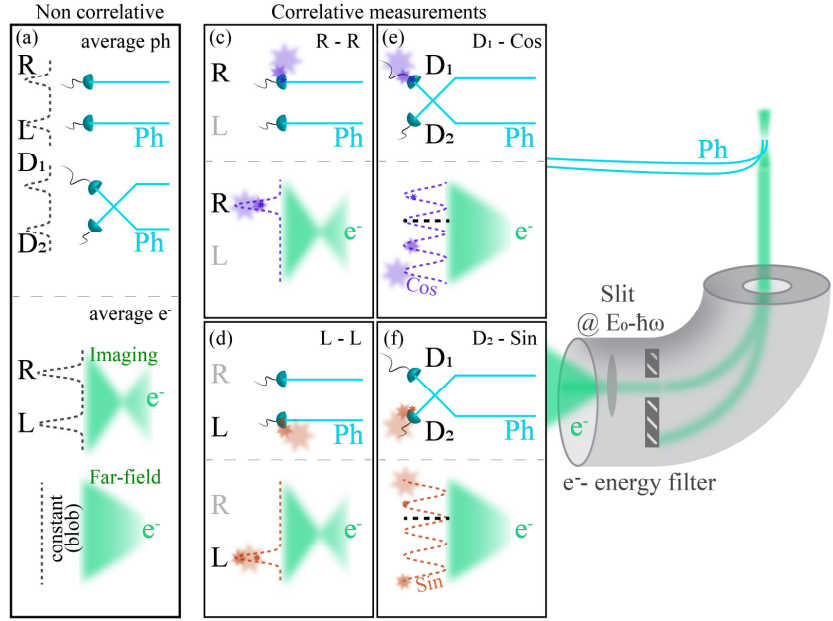

FIG. 2 – Expected electron-photon correlations. (a) uncorrelated measurements produce equal distributions of the photonic which-path (L vs. R) or relative sign ($D_1$ vs. $D_2$) phase, and of the electronic which path (imaging mode) and featureless scattering (far-field mode). (c-d) Correlative measurements would show a classical correlation between the detected paths of the photon and electron, either L&L or R&R. (e-f) The probability of an electron detection co-incident with a photon at $D_1$ or $D_2$ is a cosine or sine pattern, respectively.

In conclusion, we reviewed the phenomenon of electron-photon coupling in a novel variation of Young's experiment, albeit with a unitary interaction instead of a classical scattering or transmission through the two slits. This purely quantum scattering picture allows to avoid tracing out the contribution of energy and momentum exchanges at the two interaction regions. We first discussed a reference case of a point-particle electron coupled to either one of two photonic waveguides and the resulting photonic interference. We emphasize the effect of splitting the electron's wavefunction into two paths, each coupled to a distinguishable photonic mode. We have shown that the spatial symmetry of the electron wavefunction is not directly observed in the photon it excites, even for a pure and deterministic quantum coupling. Thus, the proposed electron-photon quantum system conforms to the complementarity of one-particle and two-particle interference, as defined by Jaeger et al. [75], where a connection between the photonic statistics and the electron wavefunction can be observed only by post selection, regardless of post-interaction manipulations of the particles. Realizing this would be the first demonstration for two distinct fundamental entities, however, as Wheeler stated, "no phenomenon is a phenomenon until it is an observed phenomenon". In the future, these qubit-like spatial states of the electron can be the backbone of quantum computation hardware or precision quantum sensing based on free electrons, that are read and undergo measurement-based operations by the photons they release.

**Acknowledgments.**

O.K. gratefully acknowledges the Young Faculty Award from the National Quantum Science and Technology program of the Israeli Planning and Budgeting Committee. H. A. gratefully acknowledges the Excellent Student Scholarship from the Center for Quantum Science and Technology at Tel Aviv University.